# Machine Learning Algorithms for Transplanting Accelerometer Observations in Future Satellite Gravimetry Missions

**Mohsen Romeshkani[1, *], Jürgen Müller[1], Sahar Ebadi[1], Alexey Kupriyanov[1], Annike Knabe[1], Nina Fletling[1], Manuel Schilling[2]**

[1] Institute of Geodesy, Leibniz University Hannover, Schneiderberg 50, Hannover 30167, Germany

[2] Institute for Satellite Geodesy and Inertial Sensing, German Aerospace Center (DLR), Callinstrase 30B, Hannover 30167, Germany

[*] Corresponding author: Mohsen Romeshkani (romeshkani@ife.uni-hannover.de)

ORCiD:

Mohsen Romeshkani (0000-0003-2774-2074)

Jürgen Müller (0000-0003-1247-9525)

Sahar Ebadi (0000-0002-4758-5566)

Alexey Kupriyanov (0000-0002-0743-5889)

Annike Knabe (0000-0002-6603-8648)

Nina Fletling (0009-0006-7727-3404)

Manuel Schilling (0000-0002-9677-0119)

**Abstract:**

Accurate and continuous monitoring of Earth's gravity field is essential for tracking mass redistribution processes linked to climate variability, hydrological cycles, and geodynamic phenomena. While the GRACE and GRACE Follow-On (GRACE-FO) missions have set the benchmark for satellite gravimetry using low-low satellite-to-satellite tracking (LL-SST), the precision of gravity field recovery still strongly depends on the quality of accelerometer (ACC) performance and the continuity of ACC data. Traditional electrostatic accelerometers (EA) face limitations that can hinder mission outcomes, prompting exploration of advanced sensor technologies and data recovery techniques. This study presents a systematic evaluation of accelerometer data transplantation using novel accelerometer configurations – including Cold Atom Interferometry (CAI) accelerometers and hybrid EA-CAI setups – and applying both analytical and machine learning-based methods. Using comprehensive closed-loop LL-SST simulations, we compare four scenarios ranging from the conventional EA-only setup to ideal dual hybrid configurations, with a particular focus on the performance of transplant-based approaches using different neural network approaches. Our results show that the dual hybrid configuration provides the most accurate gravity field retrieval. However, the transplant-based hybrid

setup – especially when supported by machine learning – emerges as a robust and cost-effective alternative, achieving comparable performance with minimal extra hardware. These findings highlight the promise of combining quantum sensor technology and data-driven transplantation for future satellite gravimetry missions, paving the way for improved global monitoring of Earth's dynamic gravity field.

## 1. Introduction

1.1. Accelerometer Data from GRACE and GRACE-FO Missions

The study of temporal and spatial variations in Earth's gravity field is vital for investigating mass movement phenomena related to climate dynamics, hydrology, and geophysical processes. The GRACE (Gravity Recovery and Climate Experiment) mission, operational from 2002 to 2017 (Tapley et al., 2004), and its successor mission GRACE Follow-On (GRACE-FO), launched in 2018 (Landerer et al., 2020), have provided a continuous record of such variations, fundamentally enhancing our ability to monitor global mass transport. Both missions utilize a Low-Low Satellite-to-Satellite Tracking (LL-SST) approach, where two spacecrafts orbit in close formation and continuously record changes in their mutual distance, primarily caused by spatial variations in Earth's gravity field.

Crucial to isolating the gravitational signals from non-gravitational perturbations – such as atmospheric drag and solar radiation pressure – are the highly sensitive accelerometers onboard each satellite. These instruments supply the data required to correct for non-gravitational accelerations, thereby allowing for more accurate gravity field reconstruction. Nevertheless, accelerometer data is susceptible to various issues, including instrument noise, drift, and periods of data loss. In order to counteract these problems, the GRACE-FO Science Data System (SDS), incorporating efforts from the Center for Space Research (CSR), NASA Jet Propulsion Laboratory (JPL), and the German Helmholtz Center for Geosciences (GFZ), has introduced robust calibration routines to improve the quality of accelerometer readings (McCullough et al., 2019).

The early failure of the accelerometer on board satellite GRACE-D of the GRACE-FO mission prompted the introduction of alternative data recovery approaches, notably the transplantation of accelerometer data from satellite GRACE-C. This approach reconstructs the missing measurements from one satellite using those from its companion, exploiting their similar orbital paths and close separation (~220 km). The transplantation method, first employed to address GRACE-B data outages during the GRACE mission (Save et al., 2006), is based on the principle that both satellites encounter nearly the same non-gravitational accelerations. By applying time and orientation adjustments, synthetic accelerometer data can be generated for the affected spacecraft.

While this strategy is effective under many conditions, it has limitations, particularly when asymmetric environmental effects, such as spatial variations in atmospheric drag or solar radiation, become significant. Such asymmetries grew more pronounced during the later years of the GRACE mission, manifesting as biases in the estimated low-degree gravity field coefficients derived from transplanted data (Loomis et al., 2020; Behzadpour et al., 2021; Bandikova et al., 2019). In addition, manoeuvre-induced accelerations, such as those resulting from thruster firings, can introduce further errors into the reconstructed datasets (Meyer et al., 2011).

1.2. Progress in Accelerometer Technologies

Despite the important advances achieved through GRACE and GRACE-FO, instrument-level challenges persist. A central issue is the noise behaviour and low-frequency drift of EA, which can degrade the long-term stability required for precise gravity field mapping. EAs are known for their low noise at moderate to high frequencies but exhibit a significant drift over extended periods, complicating efforts to estimate time-variable biases and scale factors (Kupriyanov, 2025; Knabe, 2023).

To address these shortcomings, recent research has explored new sensor concepts, with Cold Atom Interferometry (CAI) emerging as a particularly promising technology. CAI accelerometers employ ultracold atom clouds manipulated by laser pulses to detect acceleration, offering exceptional long-term stability and highly reliable scale factors linked directly to laser frequency stability. In a CAI sensor, acceleration is measured through phase shifts between atomic states after interaction with counter-propagating lasers (Lévèque et al., 2022; Antoine and Bordé, 2003). Numerical studies (Abrykosov et al., 2019; Müller and Wu, 2020; Migliaccio et al., 2023; Romeshkani et al., 2023; Zingerle et al., 2024; Romeshkani et al., 2025a; Romeshkani et al., 2025b; HosseiniArani et al., 2025) indicate that CAI instruments could significantly enhance satellite gravimetry performance.

Nevertheless, CAI accelerometers also present limitations, primarily due to their relatively long interrogation times, which can lead to unobserved short-term variations in acceleration. Hybrid sensors, combining the complementary strengths of EAs (for high-frequency measurements) and CAI (for long-term stability), are being investigated as a solution. Several studies (Zahzam et al., 2022; Zingerle et al., 2024) have analysed hybrid sensor configurations and their potential to improve the precision and robustness of gravity field measurements in space applications.

The CARIOQA (Cold Atom Rubidium Interferometer in Orbit for Quantum Accelerometry) initiative represents a major effort to advance quantum accelerometry in space, aiming to elevate the Technology Readiness Level of these sensors. Supported by the European Union, CARIOQA's primary objective is to launch a Quantum Pathfinder Mission in the 2030s. Preparatory activities include the development of a quantum accelerometer engineering model, technological maturation, and feasibility studies for future gravity missions based on quantum or hybrid sensors (Lévèque et al., 2022).

The accurate modelling and transplantation of accelerometer data are crucial for improving the performance and reliability of future satellite gravimetry missions. Given the complex, nonlinear, and temporal nature of accelerometer time series, traditional statistical methods often fall short in capturing the full dynamics inherent in these signals. In recent years, Machine Learning (ML) has emerged as a powerful paradigm for learning intricate patterns from large volumes of satellite data, enabling robust prediction, reconstruction, and cross-mission adaptation (Goodfellow et al., 2016). In this context, neural network-based architectures – such as Multi-Layer Perceptron (MLP), Long Short-Term Memory (LSTM), and Bidirectional LSTM (BiLSTM) – have demonstrated superior capabilities in extracting meaningful representations and modelling temporal dependencies, which are essential for the task of accelerometer data transplant. The following subsections provide a detailed overview of these ML approaches and their specific advantages for time series analysis in geoscientific applications.

This paper examines the use of CAI in GRACE-type gravity missions, focusing on accelerometer data transplantation. Section 2 reviews recent advances in accelerometer technologies, including electrostatic, CAI, and hybrid instruments. Section 3 addresses non-gravitational perturbations in low Earth orbit. Section 4 describes methods for accelerometer data transplantation. Section 5 presents the simulation framework for assessing the performance of a quantum accelerometer. Section 6 details ML architectures for data transplantation and comparative evaluation. Section 7 concludes the paper.

## 2. Advances in Accelerometer Technologies for LL-SST Missions

High-precision accelerometers are fundamental to the success of LL-SST missions – such as those following the GRACE concept – by enabling the differentiation of gravitational forces from other perturbing accelerations. Given the distinct but complementary capabilities of EA and CAI ACCs, emerging research proposes a hybrid solution, integrating both sensor types to maximize the accuracy and robustness of future gravimetry missions.

2.1. GRACE-FO Electrostatic Accelerometers

EAs, such as the SuperSTAR instrument developed by ONERA, form the backbone of non-gravitational force measurements in missions like GRACE-FO. However, these sensors are characterized by an intrinsic low-frequency drift, which can introduce significant errors in gravity field estimates (Christophe et al., 2015; Kupriyanov et al., 2024b; Romeshkani et al., 2025a). The SuperSTAR (EA-GFO) is the reference model for performance assessment in GRACE-FO, and its Amplitude Spectral Density (ASD) is described by (Kornfeld et al., 2019):

$$acc_M(f) = 10^{-10} \sqrt{1 + \left(\frac{f}{0.5\text{ Hz}}\right)^4 + \frac{0.005\text{ Hz}}{f}} \ \ \frac{\text{m}}{\text{s}^2 \sqrt{\text{Hz}}}. \tag{1}$$

ONERA's expertise in developing precision accelerometers for major satellite missions – including GRACE, GOCE, and GRACE-FO – has significantly advanced the field of spaceborne inertial sensing (Dalin et al., 2020). The EA-GFO is a three-axis, capacitive device placed at the satellite's center of mass (CoM). Its core comprises a 40×40×10 mm$^3$ proof mass, which is kept motionless by electrostatic forces generated by surrounding electrodes. The resulting measurements directly reflect the non-gravitational accelerations acting on the spacecraft – including atmospheric drag, solar radiation pressure, and Earth's albedo – since the sensor is located in the CoM (Touboul et al., 2004; Frommknecht, 2008; Peterseim, 2014).

The EA-GFO delivers highly sensitive linear and angular acceleration readings. Along the radial and along-track axes, its sensitivity reaches below 0.1 nm/s$^2$/Hz$^{1/2}$ within the relevant measurement spectral range, while the cross-track axis achieves 1 nm/s$^2$/Hz$^{1/2}$ (Daras & Pail, 2017). Such performance is crucial for detecting minute external forces and refining satellite orbit determination. Recent advancements in both electrostatic and optical acceleration sensing are expected to push the boundaries of precision and stability in future gravimetry missions. These developments promise improved gravity field models, with wide-reaching implications for geoscience and climate research.

2.2. CAI Accelerometers

CAI ACCs represent a new frontier in satellite inertial sensing, utilizing ultracold atom clouds as inertial references. Here, clouds of atoms are manipulated by sequences of laser pulses, acting as beam splitters and mirrors, to generate quantum interference patterns. The phase shift observed in these patterns is directly linked to the acceleration experienced by the atoms (Pereira dos Santos & Landragin, 2007; Schilling et al., 2012; Abend et al., 2020).

One of the main advantages of CAI-based ACCs is their superior long-term stability and accurately defined scale factors, which are governed by the frequency stability of the lasers. Simulation and experimental studies (Abrykosov et al., 2019; Müller & Wu, 2020; Romeshkani et al., 2025a; Romeshkani et al., 2023) have highlighted their potential for substantially improving satellite gravity recovery. Nonetheless, the performance of CAI sensors depends on several system parameters and operational conditions (Knabe et al., 2022; Lévèque et al., 2022; Beaufils et al., 2023; HosseiniArani et al., 2025). For the present study, a CAI ACC model with a representative white noise level of $10^{-10}$ m/s$^2$/Hz$^{1/2}$ (CAI 10) is considered as indicative of the expected performance.

2.3. Hybrid Accelerometers

Hybrid ACCs, which integrate EA and CAI sensors, are gaining attention as a compelling approach to overcome the individual limitations of each technology. CAI sensors provide outstanding stability and absolute calibration, but their use is challenged by operational dead times and limited dynamic range (Lévèque et al., 2022). Conversely, EAs are renowned for their proven short-term sensitivity and established reliability in space applications, but require regular calibration to counteract drift.

In hybridized inertial sensors, CAI ACC can calibrate EA scale factors and bias, while EAs deliver high-frequent, continuous data, compensating for the temporal gaps in CAI measurements, and they can also be used to resolve the fringe ambiguity of the CAI ACC. (Abrykosov et al., 2019; Knabe, 2023; Romeshkani et al., 2025a; HosseiniArani et al., 2025). This synergy yields a measurement system that capitalizes on the complementary noise properties of both sensors, offering unprecedented precision and reliability for future gravimetry missions.

In this work, hybrid ACCs are modeled by optimally merging the noise spectra of EA and CAI sensors in the frequency domain. This method exploits the unique strengths of each sensor over their respective operational frequency ranges and often achieves better results than simple time-domain combinations. The hybrid noise model is constructed by selecting an optimal cut-off frequency – determined by the intersection point of the ASDs of the two sensor types – to ensure balanced contributions from both (Romeshkani et al., 2025a; Zingerle et al., 2024). As illustrated in Figure 1, such hybridization effectively eliminates long-term drift from the EA-GFO, while maintaining its high-frequency measurement performance. The resulting hybrid system equally leverages the precision of both technologies for robust and accurate gravity field recovery.

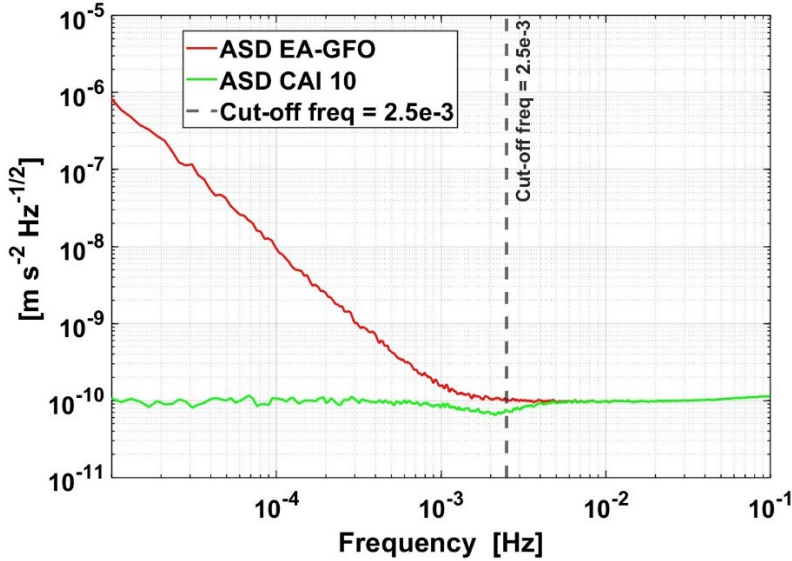

*Figure 1: ASD of hybrid accelerometer noise, EA-GFO and CAI 10*

## 3. Non-Gravitational Perturbations in Low Earth Orbit

Both GRACE and GRACE-FO operate in Low Earth Orbit (LEO), where non-gravitational forces exert a substantial influence on spacecraft trajectories. The primary contributors to these perturbations are atmospheric drag, Solar Radiation Pressure (SRP), and Earth Radiation Pressure (ERP). The magnitude and variability of these forces depend on factors such as atmospheric density, solar flux, satellite attitude, surface characteristics, and properties of the underlying Earth's surface (Klinger and Mayer-Gürr, 2016). To accurately represent these effects, satellite macro models are employed, which describe the geometry and surface properties of the spacecraft in detail (Wen et al., 2019; Wöske et al., 2019).

From all the non-gravitational disturbances, atmospheric drag is the most significant in LEOs, particularly impacting the along-track acceleration component, while SRP predominantly affects the radial component. ERP generally constitutes the smallest non-gravitational force experienced by the satellites (Klinger and Mayer-Gürr, 2016). The drag force is of particular importance for missions like GRACE, GRACE-FO and MAGIC, as it plays a crucial role in the retrieval of missing or compromised accelerometer data, which became apparent for the instruments of the satellites GRACE-B and GRACE-D (Bandikova et al., 2019; Harvey et al., 2022; Behzadpour et al., 2021; McCullough et al., 2022).

EAs, such as the EA-GFO, measure linear accelerations along three axes, but these measurements are often contaminated by high-frequency noise originating from operational events and environmental interactions. Such disturbances – arising from activities like thruster firings, structural twangs, heater operations, and magnetic torquer activations – can differ between spacecraft and must be meticulously mitigated during data processing.

Beyond linear acceleration, these instruments also provide angular acceleration measurements, which are invaluable for satellite attitude determination (Klinger and Mayer-Gürr, 2014; Sakumura et al., 2017). However, since angular accelerations are specific to each satellite's orientation and manoeuvres, these data cannot be transplanted or shared between spacecraft. Consequently, the present study focuses exclusively on the analysis and transplantation of linear acceleration measurements.

*Atmospheric Drag Modelling*

Aerodynamic drag arises from the interaction of the spacecraft with atmospheric molecules in LEO, resulting in the transfer of momentum and the creation of a net retarding force. As the dominant non-gravitational disturbance, drag significantly impacts orbit determination and gravity field recovery. To model atmospheric conditions and their influence on satellite dynamics, various empirical and physical models have been developed. Key examples include the Jacchia-Bowman 2008 model (JB08; Bowman et al., 2008), the Drag Temperature Model 2013 (DTM2013; Bruinsma, 2015; Bruinsma & Boniface 2021), and the NRLMSISE-00 model (Picone et al., 2002). These models incorporate parameters such as solar and geomagnetic activity, as well as atmospheric composition, to provide density forecasts essential for drag estimation.

Despite these advances, precise drag force modelling remains challenging due to uncertainties in satellite state vectors, attitude, surface interactions with atmospheric particles, and natural variability in atmospheric density (Moe & Moe, 2005; Prieto et al., 2014). Such uncertainties represent a major source of error in drag force predictions and, consequently, in gravity field solutions.

For dual-satellite missions like GRACE and GRACE-FO, following each other in about 25-30 s, it is generally assumed that both spacecraft experience near-identical environmental conditions. As a result, the error characteristics of the drag model are expected to be very similar for both satellites. This symmetry allows for the estimation of drag model errors using the accelerometer data from one satellite, which can then be used to reconstruct or recover missing accelerometer measurements of its companion.

## 4. Methodologies for Accelerometer Data Transplantation

Building on the similarity of environmental conditions between the GRACE satellites, this section introduces the concept of acceleration transplantation – using data from one satellite to reconstruct missing accelerometer measurements on the other.

The simplest transplantation strategy, first described by Save et al. (2006), employs only time and attitude adjustments. The time correction compensates for the lag resulting from the spatial offset between the satellites, while the attitude correction accounts for differences in their orientation with respect to the velocity vector.

Building on this approach, more advanced methods have been developed. Bandikova et al. (2019); Harvey et al., 2022 and Behzadpour et al. (2021) addressed the impact of satellite thruster firings by modelling and removing the associated residual linear accelerations. Building on this foundation, Behzadpour et al. (2021) introduced further enhancements by incorporating atmospheric drag modeling corrections and subtracting additional modeled non-gravitational forces. Their method extends the work of Bandikova et al. (2019) by including all previously applied corrections alongside these additional refinements.

At NASA's Jet Propulsion Laboratory (JPL), McCullough et al. (2022) proposed yet another strategy, which leverages filtered accelerometer data from GRACE-D as the foundation for the transplantation process. Romeshkani et al.

(2025b) were the first to demonstrate the feasibility of accelerometer data transplantation for future satellite gravimetry missions, specifically highlighting its integration with advanced CAI ACCs. Their work showed that combining transplantation techniques with CAI and hybrid accelerometer configurations can effectively overcome instrument limitations, paving the way for improved accuracy and innovative mission designs in next-generation gravity field recovery. A schematic overview of the main processing steps involved in these transplantation methodologies is illustrated in Figure 2.

(Bandikova et al., 2019 and Harvey et al., 2022)

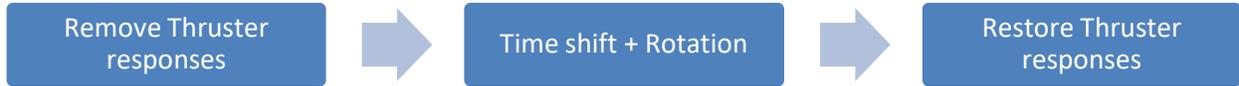

(Behzadpour et al., 2021)

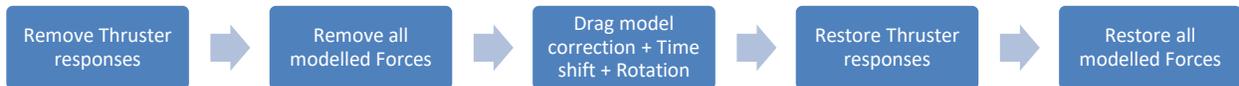

(McCullough et al., 2022)

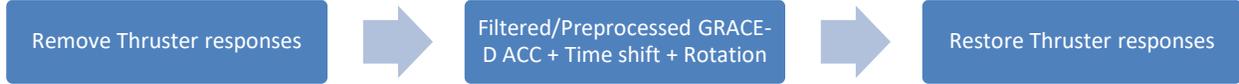

(Romeshkani et al., 2025b)

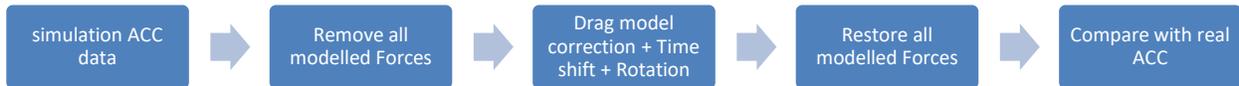

Figure 2: Four approaches for transplant accelerometer observations

## 5. Simulation Framework for Assessing Quantum Accelerometer Performance

To systematically evaluate the influence of quantum accelerometer technologies on future satellite gravimetry missions, a dedicated simulation environment has been established. The simulated scenarios are configured in a way to closely replicate the orbital and measurement characteristics of current single-pair missions, such as GRACE-FO, thereby ensuring that the analysis reflects realistic mission conditions for gravity field retrieval.

Orbital Configuration

Gravity Field Recovery (GFR) simulations were conducted using the high-performance computing cluster at Leibniz University Hannover (LUIS, 2024) to evaluate the effectiveness of the developed algorithms and numerical routines. The simulation procedure closely follows the closed-loop approach described in Romeshkani et al. (2025b) and Kupriyanov et al. (2024a). Initially, GRACE-like satellite orbits with an altitude of 478.48 km were generated using the Matlab/Simulink toolbox eXtended High Performance Satellite Dynamics Simulator (XHPS) (Wöske et al., 2016; Wöske, 2021). In the detailed orbital dynamics, the effect of the non-gravitational forces, third bodies, solid Earth tides, etc. were included (Kupriyanov, 2025). Then, the orbit simulated in XHPS was exported into the GFR software. In all simulated scenarios, a single satellite pair is assumed to operate in a near-polar orbit with an inclination of 89°, similar to the original GRACE mission parameters.

Forward Gravity Field Modeling

The forward modeling component utilizes the static gravity field model GOCO05s (Mayer-Gürr et al., 2015), incorporating gravitational signals up to spherical harmonic degree and order 120. By restricting the analysis to static gravity field contributions, the framework isolates the influence of instrument noise, thereby enabling a direct assessment of quantum ACC performance under idealized circumstances where temporal aliasing is not present. Temporal aliasing, which arises from unresolved rapid changes in the atmosphere and ocean signals, is intentionally excluded to highlight instrument-related effects.

Gravity Field Retrieval via Inverse Modeling

Gravity field estimation is performed through a least-squares adjustment, using static spherical harmonics as basic functions. The LL-SST observations are treated as range-rate measurements, processed according to an integral equation framework with short-arc methodology (Mayer-Gürr, 2006). The stochastic characteristics of the observations are represented by covariance matrices derived from the spectral densities of the simulated instrument noise. A retrieval period of one month is adopted, and the maximum spherical harmonic degree and order is set to 120, consistent with the standards of GRACE-type mission data analysis.

Figure 3 presents an overview of the computational workflow for the LL-SST method. The process initiates with the simulation of a true gravity field alongside a reference model, followed by the calculation of differential gravitational and non-gravitational accelerations between the two satellites. Simulated instrument noise is then added to these signals, leading to the formulation of a system of equations for subsequent gravity field recovery.

Evaluation Objectives

By employing this comprehensive and structured simulation framework, the study aims to deliver a robust assessment of the advantages that quantum-based accelerometers may offer for satellite gravimetry. The results provide key insights into the potential improvements such technology could bring to future missions dedicated to monitoring Earth's gravity field.

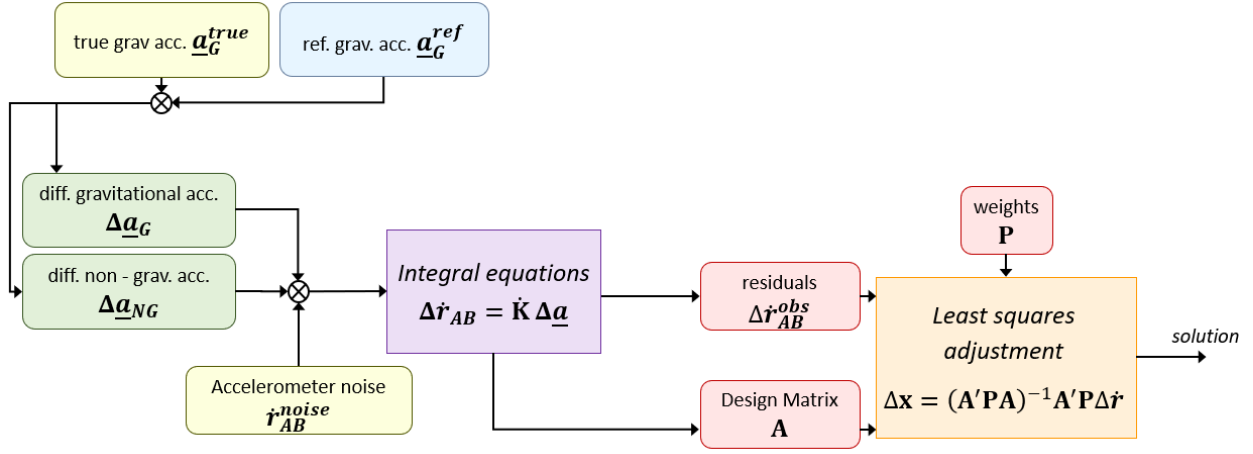

*Figure 3: Closed-loop simulation in the LL-SST concept (Romeshkani et al., 2025b)*

## 6. Machine Learning Architectures for Accelerometer Data Transplant

Selecting an appropriate ML architecture is critical for effectively modeling accelerometer data, which are inherently sequential and often exhibit complex temporal dependencies. Among the various approaches available, feed-forward neural networks such as the Multi-Layer Perceptron (MLP) can capture nonlinear relationships in fixed-length data segments but do not explicitly address time order. In contrast, Recurrent Neural Networks (RNN) like the Long Short-Term Memory (LSTM) are designed to model sequences by maintaining memory of previous observations, enabling the learning of long-term temporal dynamics. Further extending this capability, Bidirectional LSTM (BiLSTM) networks process data in both forward and backward directions, thereby incorporating information from past as well as future time steps. Thanks to these architectures, it becomes possible to transplant, reconstruct, and predict ACC readings in satellite applications more accurately, taking full advantage of the underlying structure of time series data (Graves & Schmidhuber, 2005; Hochreiter & Schmidhuber, 1997; Goodfellow et al., 2016).

6.1. Multi-Layer Perceptron

In this subsection, we present an overview of the MLP architecture and its application to satellite accelerometer data modeling. MLPs are classical feed-forward artificial neural networks, capable of approximating complex nonlinear mappings between input and output data. Although MLPs lack explicit mechanisms for modeling temporal dependencies, they can be effective when provided with appropriately constructed feature vectors that include temporal context (Chen et al., 2023; Tang and Zhang., 2024; Das et al., 2023).

**Definition 1**

Let us denote the input samples as $S = \{a^{(i)}\}_{i=1}^{N}$ and the corresponding targets as $T = \{t^{(i)}\}_{i=1}^{N}$ where each input vector $a^{(i)} \in R^d$ is a fixed-length feature representation (for instance, a flattened window of accelerometer values), and $t^{(i)} \in R$ is the associated target value. Here, $N$ denotes the number of training examples, and $d$ the dimensionality of the input feature space (Goodfellow et al., 2016; Bishop, 2006).

The MLP comprises $M$ layers, with each layer $m$ is defined by a weight matrix $\mathbf{\Theta}_m$ and a bias vector $\boldsymbol{\beta}_m$. The activations through the network are computed as follows (Rumelhart et al., 1986; Goodfellow et al., 2016):

$$\mathbf{z}_0^{(i)} = \mathbf{a}^{(i)}, \tag{2}$$

$$\mathbf{z}_m^{(i)} = \psi_m(\mathbf{\Theta}_m \mathbf{z}_{m-1}^{(i)} + \boldsymbol{\beta}_m), \quad m = 1, 2, \dots, M-1, \tag{3}$$

$$\hat{t}^{(i)} = \mathbf{\Theta}_M \mathbf{z}_{M-1}^{(i)} + \boldsymbol{\beta}_M, \tag{4}$$

where $\psi_m(.)$ is the activation function for layer $m$ (commonly ReLU, sigmoid, or hyperbolic tangent), and $\hat{t}^{(i)}$ denotes the model's prediction for the i-th input. An MLP diagram should illustrate the flow of data from the input vector $\mathbf{a}^{(i)}$ through two or more hidden layers (with activations $\psi$), culminating in the output node $\hat{t}^{(i)}$. Each connection represents a learned weight, with biases applied at each neuron.

**Definition 2**

The model parameters $P = \{\mathbf{\Theta}_m, \boldsymbol{\beta}_m\}_{m=1}^{M}$ are trained to minimize the discrepancy measure between predictions and true targets. A typical objective is the Mean Squared Error (MSE) or Mean Absolute Error (MAE):

$$J_{MAE}(P) = \frac{1}{N} \sum_{i=1}^{N} |\hat{t}^{(i)} - t^{(i)}| \tag{5}$$

or

$$J_{MSE}(P) = \frac{1}{N} \sum_{i=1}^{N} (\hat{t}^{(i)} - t^{(i)})^2. \tag{6}$$

Optimization is typically carried out using gradient-based algorithms such as stochastic gradient descent (SGD) or the Adam optimizer (Goodfellow et al., 2016).

The universal approximation theorem shows that MLPs with a single hidden layer can approximate any continuous function on compact subsets of $R^d$ (Hornik et al., 1989), which underlies their success in a wide range of applications, including geoscientific data modelling and prediction (LeCun et al., 2015).

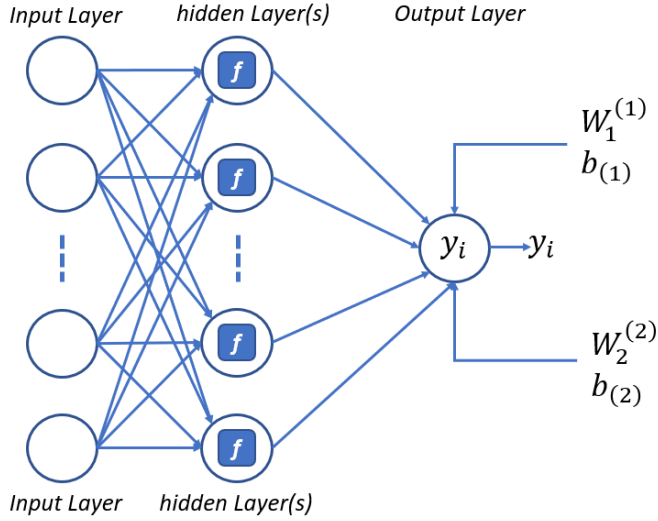

*Figure 4: Architecture of a feed-forward neural network MLP. The network consists of an input layer, one hidden layer with nonlinear activation functions f, and an output layer producing prediction y. Each layer is fully connected, with learned weights W$^{(1)}$, W$^{(2)}$ and biases b$_{(1)}$, b$_{(2)}$.*

6.2. LSTM Neural Networks

LSTM networks are a specialized type of RNN designed to address the challenges of learning long-range dependencies in sequential data (Hochreiter & Schmidhuber, 1997; Greff et al., 2017). Their unique gating mechanisms allow them to overcome the vanishing and exploding gradient problems that commonly affect traditional RNNs, enabling stable learning from time series with complex, long-term patterns (Goodfellow et al., 2016; Das et al., 2023; ).

6.2.1. Formal Description

Suppose we have an input sequence $U = [\mathbf{x}_1, \mathbf{x}_2, \ldots, \mathbf{x}_T]$ where each input vector $\mathbf{x}_t \in R^p$ is observed at time $t = 1, 2, \ldots, T$ and a corresponding output sequence $B = [\boldsymbol{v}_1, \boldsymbol{v}_2, \ldots, \boldsymbol{v}_T]$ with $v_t \in R^q$. Here, $p$ denotes the number of features in each input vector $\mathbf{x}_t$, and $q$ denotes the number of output values predicted at each time step in $v_t$. The LSTM unit maintains two types of internal state at each time step: a hidden state $\mathbf{h}_t$ and a cell state $\mathbf{C}_t$. The cell state serves as a persistent memory, while the hidden state provides output to the next layer or time step (Gou et al., 2023).

At each time step $t$, the LSTM cell computes

$$\mathbf{f}_t = \sigma(\mathbf{W}_f \mathbf{x}_t + \mathbf{U}_f \mathbf{h}_{t-1} + \mathbf{b}_f) \qquad (forget\ gate) \qquad (7)$$

$$\mathbf{i}_t = \sigma(\mathbf{W}_i \mathbf{x}_t + \mathbf{U}_i \mathbf{h}_{t-1} + \mathbf{b}_i) \qquad (input\ gate) \qquad (8)$$

$$\mathbf{o}_t = \sigma(\mathbf{W}_o \mathbf{x}_t + \mathbf{U}_o \mathbf{h}_{t-1} + \mathbf{b}_o) \qquad (output\ gate) \qquad (9)$$

$$\tilde{\mathbf{c}}_t = tanh(\mathbf{W}_c \mathbf{x}_t + \mathbf{U}_c \mathbf{h}_{t-1} + \mathbf{b}_c) \quad (cell\ candidate) \quad (10)$$

$$\mathbf{c}_t = \mathbf{f}_t \odot \mathbf{c}_{t-1} + \mathbf{i}_t \odot \tilde{\mathbf{c}}_t \quad (cell\ state\ update) \quad (11)$$

$$\mathbf{h}_t = \mathbf{o}_t \odot tanh(\mathbf{c}_t) \quad (hidden\ state) \quad (12)$$

where

$\sigma(\cdot)$ is the sigmoid activation,

$tanh(\cdot)$ is the hyperbolic tangent,

$\odot$ denotes element-wise multiplication,

$\mathbf{W}_*, and\ \mathbf{U}_*\ and\ \mathbf{b}_*$ are trainable weight matrices and bias vectors.

The gates ($\mathbf{f}_t, \mathbf{i}_t, \mathbf{o}_t$) regulate the flow of information:

- The forget gate $\mathbf{f}_t$ decides what fraction of the previous cell state to retain.
- The input gate $\mathbf{i}_t$ controls how much new information from the candidate state $\tilde{\mathbf{c}}_t$ is added.
- The output gate $\mathbf{o}_t$ determines how much of the cell state contributes to the hidden state and output.

6.2.2. Output Mapping and Training

For sequence regression or forecasting, the final hidden state ($\mathbf{h}_T$) or a sequence of hidden states ($\{\mathbf{h}_T\}_{t=1}^T$) can be passed to a dense layer to yield the network output:

$$\hat{v}_T = \mathbf{W}_{out} \mathbf{h}_T + \mathbf{b}_{out} \quad (13)$$

where $\mathbf{W}_{out}$ and $\mathbf{b}_{out}$ are output weights and biases.

The model parameters (all weights and biases) are trained by minimizing a loss function such as MSE between predicted output $\hat{v}_T$ and true output $v_T$ over the training set, typically using the Adam or RMSprop optimizer (Kingma & Ba, 2015).

6.2.3. Practical Remarks and Applications

LSTM networks are widely recognized for their capacity to model nonlinear, long-term dependencies in sequential geoscientific data, including gravity field recovery, satellite signal prediction, and accelerometer data modeling. Their architecture enables effective learning from noisy and non-stationary time series, a common characteristic of satellite-based measurements.

In summary, the LSTM's architectural innovations – gating mechanisms, persistent memory, and stable gradient flow – make it a standard tool for time series modeling in ML and geoscience domains.

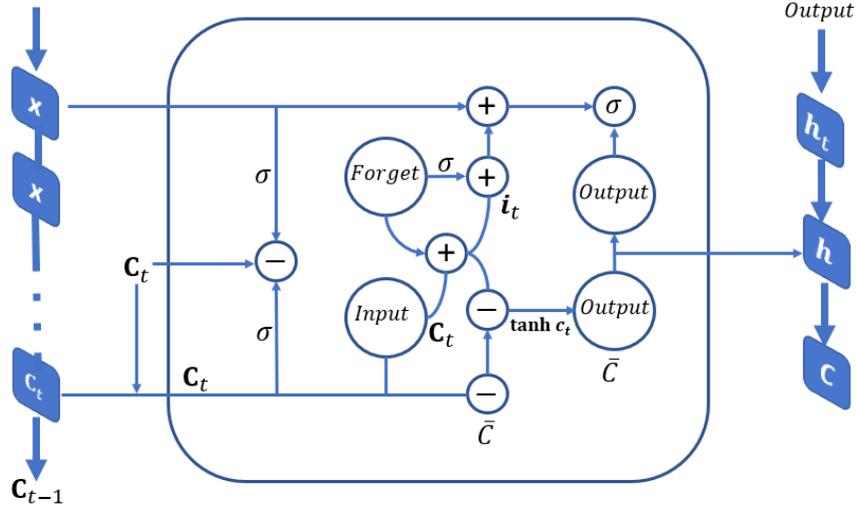

*Figure 5: Structure of a Long Short-Term Memory (LSTM) cell. The LSTM processes an input sequence $x_1, x_2, \ldots, x_n$, along with the previous cell state $\mathbf{C}_{t-1}$, to compute the current cell state $\mathbf{C}_t$ and hidden state (output) $\mathbf{h}_t$. The flow of information is controlled by three gates: The **Forget gate** (with activation $\sigma$), the **Input gate** (which generates candidate cell state $\tilde{\mathbf{c}}_t$), and the **Output gate** (also using $\sigma$ and $\tanh$ activations). Intermediate variables include $\mathbf{i}_t$ (input gate activation), and the updated outputs are passed forward through $\mathbf{C}_t$ and $\mathbf{h}_t$, preserving both long-term and short-term dependencies.*

6.3. BiLSTM Networks

BiLSTM networks extend the standard LSTM architecture by incorporating two separate LSTM layers that process the input sequence in opposite directions: one from the start to the end (forward) and the other from the end to the start (backward) (Graves & Schmidhuber, 2005; Goodfellow et al., 2016). This allows the network to access both past and future context at each time step, resulting in richer sequence representations (Fan et al., 2023).

6.3.1. Formal Definition and Architecture

Suppose the input sequence is denoted as $U = [\mathbf{x}_1, \mathbf{x}_2, \ldots, \mathbf{x}_T]$, where each $\mathbf{x}_t \in R^p$ and $T$ is the sequence length. The BiLSTM consists of:

- A **forward LSTM** that computes a hidden state $\mathbf{h}_t^{fwd}$ for each time step $t$ by traversing the sequence from $t = 1$ to $t = T$.
- A **backward LSTM** that computes a hidden state $\mathbf{h}_t^{bwd}$ for each time step $t$ by traversing the sequence from $t = T$ to $t = 1$.

At each time step $t$, the BiLSTM concatenates the two hidden states:

$$\mathbf{h}_t^{Bi} = \begin{bmatrix} \mathbf{h}_t^{fwd} \\ \mathbf{h}_t^{bwd} \end{bmatrix}. \tag{14}$$

The output for time step $t$ is then typically computed by applying a dense (fully connected) layer to $\mathbf{h}_t^{Bi}$.

$$\hat{y}_t = \boldsymbol{W}_{\text{out}} \mathbf{h}_t^{Bi} + \boldsymbol{b}_{out} \tag{15}$$

Where $\boldsymbol{W}_{\text{out}}$ and $\boldsymbol{b}_{out}$ are the output weight matrix and bias.

Each LSTM (forward and backward) uses the standard LSTM equations to compute its respective hidden and cell states (Hochreiter & Schmidhuber, 1997).

6.3.2. Advantages and Applications

By utilizing both previous and future information, BiLSTM models are especially effective for time series analysis where predictions can benefit from the full temporal context (Schuster & Paliwal, 1997). In geoscientific applications, such as satellite accelerometer data analysis and environmental time series forecasting, BiLSTMs have demonstrated improved performance over unidirectional models (Fan et al., 2023).BiLSTM networks are widely used in sequence labeling, event detection, signal denoising, and other tasks where both past and future data points are informative (Mousavi et al., 2020; Kratzert et al., 2019).

6.3.3. Training

All network parameters, including those of both the forward and backward LSTMs and the output layer, are optimized jointly to minimize a loss function such as the MSE or MAE. Training is commonly performed with optimizers like Adam (Kingma & Ba, 2015).

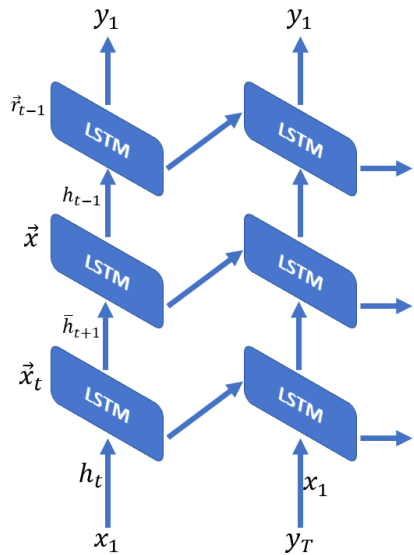

*Figure 6: BiLSTM architecture. Two LSTM layers process the input sequence in forward and backward directions. At each time step, their hidden states are concatenated to form a context-aware representation. In the diagram, $h_t$ and $\vec{r}_{t-1}$ represent hidden states from the forward and backward passes, respectively, and $\bar{h}_{t+1}$ denotes the hidden state passed from time step $t+1$ to $t$ in the backward LSTM.*

### 6.4. Features and Target Definition

In this study, ML techniques (MLP, LSTM, and BiLSTM) are applied not in the conventional context of time series forecasting or classification, but rather as tools for transplanting ACC observations between GRACE-like satellites. Unlike standard ML workflows – which typically involve splitting the dataset into training and validation subsets (e.g., 80% training and 20% testing) – our application focuses on learning transferable representations across spatial and temporal domains for a specific, physically-informed reconstruction task. The primary objective is not to predict future values in a sequence, but to model the relationship between the ACC data at one satellite (or orbit arc) and its counterpart, using learned inter-satellite patterns. As such, the training and evaluation strategy has been adapted to this unique use case, and care has been taken to avoid data leakage and ensure meaningful comparisons.

In the context of ML-based accelerometer data transplant, careful selection and definition of features and target variables are essential for model performance and interpretability. For this study, the features correspond to the set of input variables provided to the neural network at each time step, while the target represents the desired output that the model aims to reconstruct.

#### 6.4.1. Feature Construction

Each satellite, designated as SAT-C and SAT-D, is subject to a total acceleration signal that consists of the following components:

**Modeled acceleration ($ACC_C^M$, $ACC_D^M$)**: This acceleration includes contributions derived from non-gravitational force models, encompassing effects such as atmospheric drag and SRP.

**Accelerometer noise ($ACC^N$)**: It encompasses the instrumental noise intrinsic to the onboard accelerometer.

**Unmodeled acceleration ($ACC^R$)**: Residual non-gravitational forces that are not captured by the existing models – such as errors due to mismodeling atmospheric density variations – also contribute to the total acceleration. To quantify these unmodeled components, the difference between two atmospheric drag models, JB08 and NRLMSISE-00 (NRLM), is employed as an estimate. Accordingly, for each satellite, the total acceleration can be expressed as

$$ACC^{Total} = ACC^M + ACC^N + ACC^R .  \quad (16)$$

For machine learning approaches to accelerometer data transplantation, a range of input features is constructed from different sources and models. Specifically, the primary features utilized in this study include synthetic accelerations, denoted as $ACC_C^M$ and $ACC_D^M$, which are generated using the JB08 atmospheric drag model for satellites SAT-C and SAT-D. These features represent the model-derived non-gravitational accelerations and provide the ML algorithm with an approximation of the underlying forces acting on both spacecraft. Additionally, another set of features is derived from the NRLM atmospheric model, yielding $ACC_C^{Total}$ and $ACC_D^{Total}$. These quantities are treated as proxies for "real" observations, as the NRLM model offers an independent estimate of the total non-gravitational acceleration experienced by each satellite. Incorporating data from both JB08 and NRLM models allows the ML model to learn relationships and discrepancies between different atmospheric density representations and their effect on acceleration signals.

The ultimate objective in this ML-based framework is to reconstruct the accelerometer observation at SAT-D under the scenario where a CAI accelerometer is assumed to be operating on that satellite. In other words, the ML model is trained to estimate the high-fidelity acceleration measurement that would be observed at SAT-D with a CAI sensor, using both the synthetic and model-based acceleration features from both satellites as input. This strategy leverages the strengths of data-driven methods to address potential gaps or errors in traditional model-based approaches and to optimally combine multiple sources of information for improved accelerometer data transplantation and gravity field recovery.

## 7. Evaluation of Accelerometer Configurations for Future Satellite Gravimetry Missions

To systematically investigate the effect of advanced accelerometer technologies and data transplantation strategies on gravity field recovery, we consider four distinct instrument configurations representative for potential future satellite gravimetry missions. These scenarios, inspired by the GRACE and GRACE-FO mission designs, encompass both conventional EA and quantum-based CAI accelerometers. The selected configurations are summarized in the following.

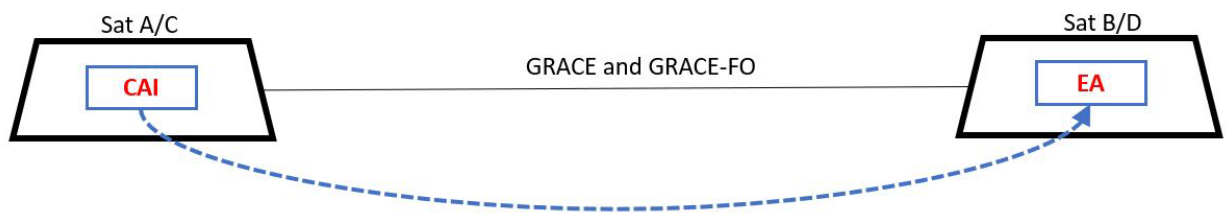

*Figure 7: Transplant of CAI observation (Romeshkani et al., 2025b)*

In our previous work (Romeshkani et al., 2025b), we systematically compared four distinct accelerometer configurations for satellite gravimetry missions to assess the impact of emerging quantum sensor technologies and data transplantation strategies. These scenarios included: (1) a conventional baseline with both satellites equipped with EAs; (2) a mixed configuration with a conventional EA on the leading satellite and a CAI accelerometer on the follower; (3) a fully redundant system where both satellites carried both EA and CAI sensors, maximizing measurement precision and redundancy; and (4) a transplantation-based hybrid approach, in which CAI and EA data were exchanged between the satellites to create a virtual hybrid dataset without requiring additional instrumentation. Comparative analysis of gravity field retrieval across these configurations provided critical insight into the advantages and trade-offs of quantum and hybrid accelerometer architectures for future gravimetry missions, offering guidance for the design of next-generation satellite systems.

Building upon these findings, the present study focuses specifically on the fourth scenario – the transplantation-based hybrid configuration. In our previous work, we explored this approach using a fully analytical framework. However, we now seek to enhance and generalize the transplantation process by introducing the three distinct ML techniques MLP, LSTM, and BiLSTM. As described in the earlier sections, each of these ML models offers unique strengths for capturing the temporal and nonlinear relationships inherent in satellite accelerometer data.

A key limitation identified in the analytical transplantation approach was its reduced effectiveness (drift in low frequency part) when transplanting CAI accelerometer observations from one satellite to another, whereas the transplantation of EA observations proved to be straightforward and reliable. Accordingly, the focus of this study is on improving the transplantation of CAI-derived acceleration data using advanced ML techniques. We restrict our analysis to the CAI-to-satellite transplantation problem, systematically evaluating the potential of MLP, LSTM, and BiLSTM models to overcome the limitations of analytical methods and improve the accuracy and robustness of CAI data recovery in future satellite gravimetry missions.

Figure 8 presents a comparative analysis of the three ML approaches – MLP, LSTM, and BiLSTM – against the conventional analytical transplantation method. This figure shows the ASDs of noise in CAI accelerometer observations after hybridization by transplantation with EA data at the destination satellite. For reference, the ideal scenario is given by the hybrid case, where both CAI and EA instruments are physically present on the two satellites, yielding optimal noise performance.

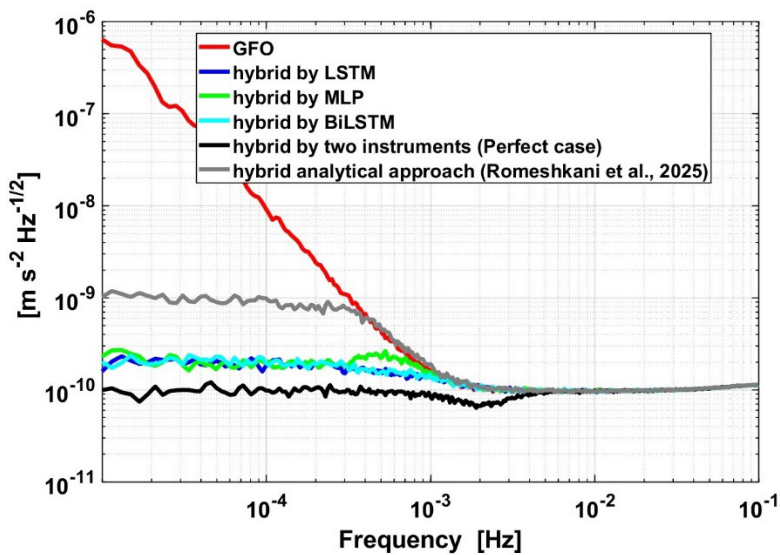

*Figure 8: Comparative ASD noise of the hybrid transplant cases*

The results clearly illustrate that all three ML-based transplantation approaches substantially mitigate the low-frequency drift in the CAI noise spectrum when compared to the analytical approach. Notably, both the LSTM and BiLSTM models exhibit nearly identical performance, indicating strong agreement between these two RNN architectures in modelling temporal dependencies in the data. The MLP approach, while also effective in suppressing low-frequency drift, exhibits particularly low noise in the medium-frequency range, although it performs slightly worse than the LSTM and BiLSTM models in this band.

The MLP approach, while effective in suppressing low-frequency drift, exhibits higher noise levels in the medium-frequency range compared to the LSTM and BiLSTM models, indicating inferior performance in this band.

These findings highlight the potential of ML-based techniques to approximate the performance of an actual onboard hybrid sensor configuration, offering significant improvements in noise characteristics – especially in the challenging low-frequency regime. Overall, the close alignment between LSTM and BiLSTM results, combined with the slightly reduced performance of MLP in the medium-frequency range, highlights the complementary strengths and trade-offs of these ML strategies for future satellite gravimetry data processing.

7.1. Comparative analysis at the level of gravity field recovery

Figure 9 offers a detailed comparison of gravity field recovery capabilities across several satellite instrument configurations and processing strategies, with a particular emphasis on the impact of both analytical and ML-based transplant approaches. The error in Equivalent Water Height (EWH) is plotted as a function of spherical harmonic degree, serving as a comprehensive metric for evaluating each scenario's ability to resolve spatial details in the time-variable gravity field.

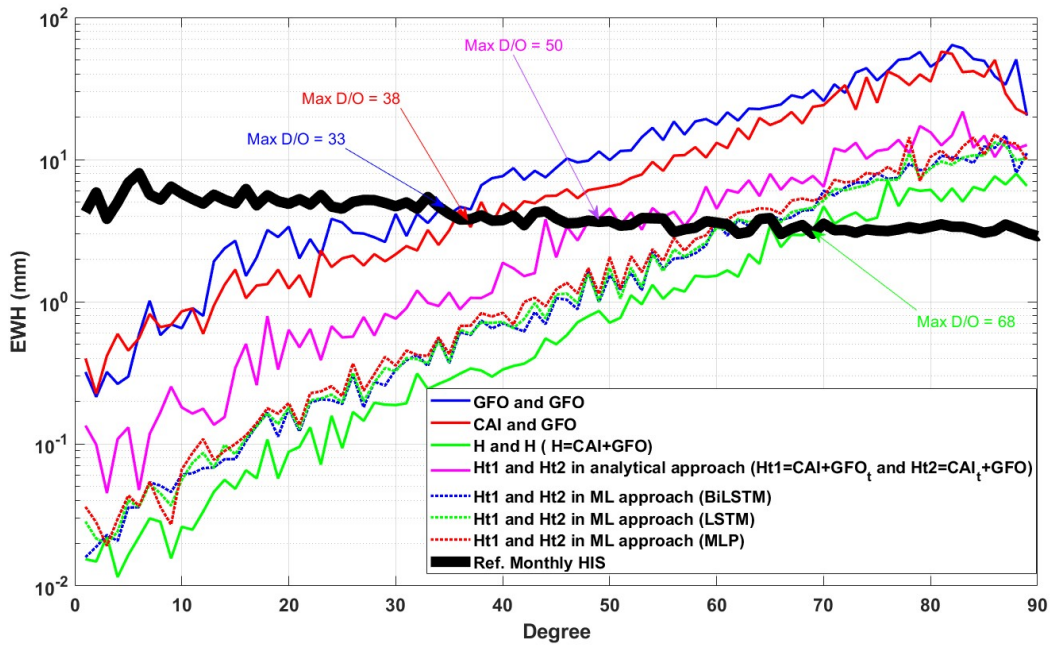

*Figure 9: Comparison between seven cases in recovering the time-variable Earth's gravity field using ML approaches w.r.t. the monthly reference mean signal (ESM components HIS)*

**Scenario 1** (Case 1 GFO_GFO, solid blue curve) represents the conventional baseline, with both satellites equipped with EAs. This case is characterized by relatively high EWH errors, especially at higher spherical harmonic degrees. The limitation is most evident in the maximum resolvable Degree/Order (D/O = 33), reflecting the inherent constraints of relying solely on classical EA technology for non-gravitational force measurement.

**Scenario 2** (Case 2 CAI_GFO, solid red curve) explores the effect of integrating quantum sensor technology by outfitting the trailing satellite with a CAI accelerometer, while the leading satellite retains a conventional EA. This hybrid pairing reduces the overall error and extends the maximum reliable recovery to D/O = 38. The improvement demonstrates the advantage of quantum accelerometry in reducing errors associated with non-gravitational force modeling, even when deployed on just one of the two satellites.

**Scenario 3** (Case 3 H_H, solid green curve) represents the idealized dual hybrid configuration, where both satellites carry both an EA and a CAI ACC, providing a total of four independent accelerometer data streams. This setup achieves the best performance among all cases, with respect to the monthly reference mean signal (HIS) and a maximum D/O of 68. The synergy between CAI and EA enables superior suppression of noise across the frequency spectrum, making this configuration the benchmark for future gravimetry missions—albeit with increased system complexity and cost due to instrument redundancy.

**Scenario 4** (Case 4 Analytical transplant approach, solid magenta curve) investigates a transplant-based hybrid configuration, with additional focus on the processing approaches. In this scenario, the leading satellite is equipped with an EA, while the follower carries a CAI accelerometer. The configuration **Ht1** represents a hybrid case that combines CAI ACC observations from the follower spacecraft with EA observations transplanted from the leader. Conversely, **Ht2** is a hybrid case where CAI ACC data are transplanted from the follower to the leader and then combined with the leader's EA observations. The analytical transplantation of CAI and EA data significantly reduces the error compared to Scenarios 1 and 2, allowing accurate recovery up to D/O = 50 (Romeshkani et al., 2025b). However, the analytical method remains less effective at fully mitigating the low-frequency drift and transplant process noise w.r.t. Fig 8.

The introduction of ML methods for transplantation instead of the analytical approach, in particular BiLSTM, LSTM, and MLP (dotted blue, green, and red lines, respectively) provides a marked enhancement in performance. All three ML approaches enable the transplant-based configuration to closely approach the accuracy of the dual hybrid case, with reliable recovery up to D/O = 61. The BiLSTM and LSTM results are nearly indistinguishable, indicating strong model robustness and consistency in learning temporal dependencies inherent in accelerometer data. The MLP approach also contributes to reducing drift but performs slightly worse than the RNN models in the mid-frequency regime (Fig. 8).

Figure 10 presents the geographical distribution of errors in EWH, along with the spherical harmonic error spectra of coefficient differences for each scenario. To ensure consistency and facilitate comparison, the colour bar limits for spatial residuals are uniformly set to ±0.1 m EWH for all global maps. The spherical harmonic error spectra of coefficient differences, plotted in logarithmic scale for each scenario, are shown in Figure 11. The results, both in the spatial domain and in the two-dimensional spherical harmonic error spectra, align well with Figure 9. Notably, Case 3 H_H (subplot c) exhibits the lowest residual magnitudes, which are also more isotropically distributed. In contrast, Case 1 GFO_GFO and Case 2 CAI_GFO (subplots a and b) display more pronounced directional patterns in their residuals, particularly appearing as stripes in the North–South direction.

The expanded comparative analysis reveals several important insights. The dual hybrid configuration (Scenario 3), in which both satellites are equipped with both CAI and EA accelerometers, establishes the upper benchmark for gravity field retrieval accuracy, though it does so with increased hardware complexity and cost. Introducing a single CAI ACC (Scenario 2) already leads to substantial improvements over the conventional EA-only baseline (Scenario 1), highlighting the clear advantage of quantum sensor technology for future satellite missions. The analytical transplant approach in Scenario 4 significantly narrows the performance gap between the mixed and ideal cases, yet it remains constrained by transplant process noise and residual modelling errors. Notably, the ML-based transplantation methods (Scenario 4, dotted lines in Figure 9) further close this gap, in some spectral ranges nearly matching the performance of the dual hybrid setup, while preserving a streamlined instrument configuration. In summary, these results demonstrate that, while the dual hybrid scenario remains technically optimal, ML-based transplant approaches offer a compelling and practical path forward – achieving high-precision gravity field estimates with fewer onboard sensors, thus enhancing both the scientific return and cost-effectiveness of future satellite gravimetry missions.

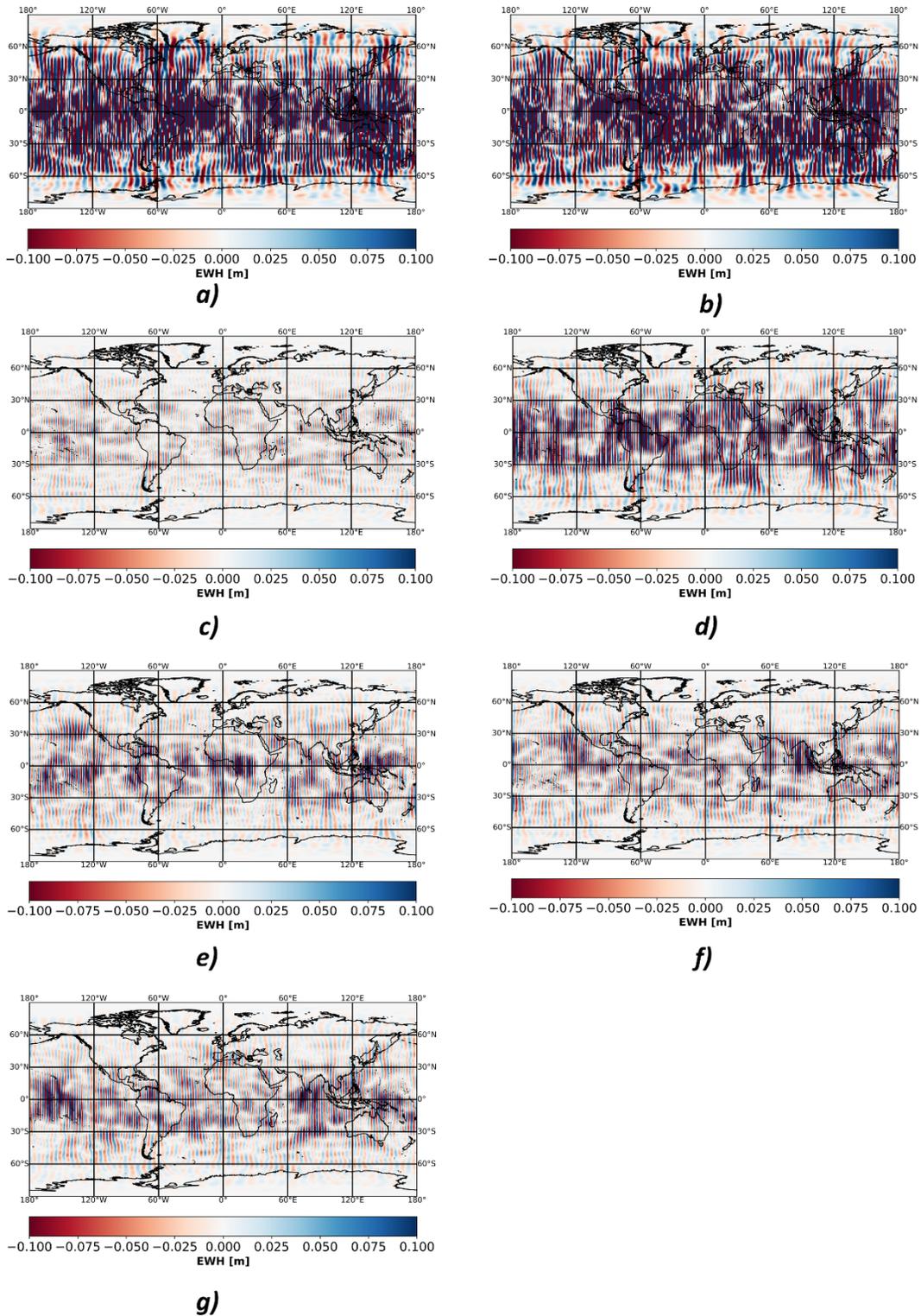

*Figure 10: Residuals of the recovered gravity models, plotted on global maps in EWH w.r.t. GOCO05s. a) – Case 1 GFO_GFO; b) – Case 2 CAI_GFO; c) – Case 3 H_H; d) – Case4 Analytical approach; e) – Case 4 MLP; f) – Case4 LSTM; g) – Case 4 BiLSTM.*

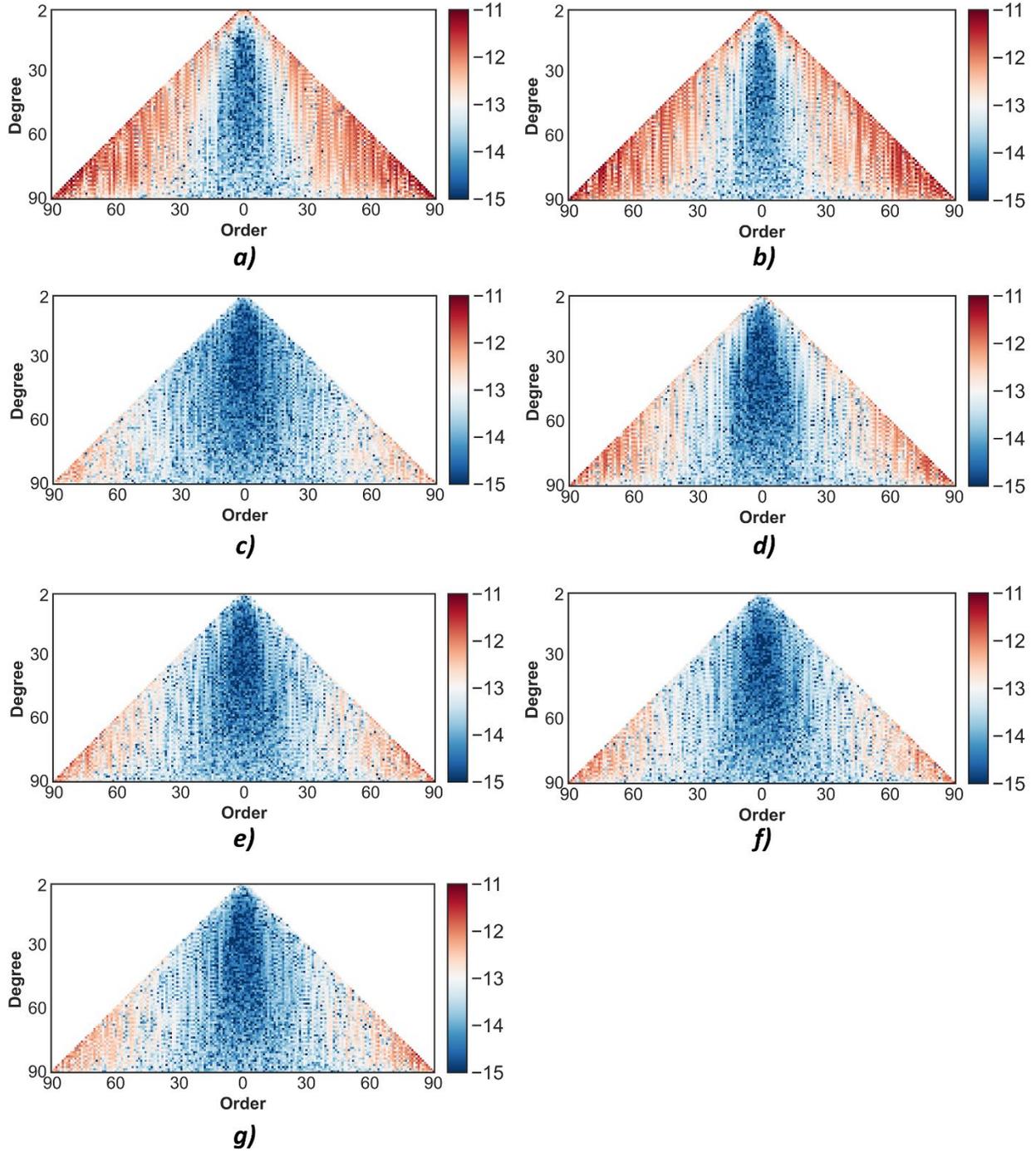

*Figure 11: 2D spherical harmonic error spectra of the true errors w.r.t. GOCO05s in logarithm scale. a) – Case 1 GFO_GFO; b) – Case 2 CAI_GFO; c) – Case 3 H_H; d) – Case4 Analytical approach; e) – Case 4 MLP; f) – Case4 LSTM; g) – Case 4 BiLSTM.*

## 8. Conclusions

This study offers a comprehensive assessment of accelerometer data transplantation strategies and sensor configurations for enhancing gravity field recovery in future satellite gravimetry missions. By systematically modeling real and synthetic accelerations, noise components, and their propagation between spacecraft, we evaluate the relative merits of various setups – including conventional EA, quantum-based CAI ACC, and hybrid configurations – within GRACE-like mission architectures.

Our findings reaffirm that the dual hybrid configuration, where both satellites are equipped with both EA and CAI ACCs, achieves the highest accuracy in gravity field estimation. This arrangement effectively leverages the complementary strengths of each sensor type, minimizing errors across all spectral ranges. However, this optimal performance is offset by increased hardware complexity and mission cost.

The transplant-based hybrid configuration emerges as a highly promising and practical alternative. By exchanging accelerometer observations between the satellites, this approach delivers a substantial improvement in gravity field recovery over single-CAI or conventional setups, despite the introduction of some transplant error. Notably, our study demonstrates that applying advanced ML methods – specifically MLP, LSTM, and BiLSTM neural networks – to the transplantation of CAI data can significantly reduce transplant noise and approach the accuracy of the ideal dual hybrid scenario. These ML-based approaches enable robust, high-resolution gravity field retrieval while maintaining a streamlined, cost-effective instrument suite.

In summary, this research highlights the transformative potential of quantum accelerometry and data-driven transplantation techniques for next-generation satellite gravimetry missions. By balancing precision, redundancy, and resource efficiency, these innovations pave the way for improved global monitoring of Earth's time-variable gravity field, with far-reaching benefits for understanding climate dynamics, water resources, and geophysical processes.

## Declarations

### Funding

The investigations were supported by funding of the Deutsche Forschungsgemeinschaft (DFG) –TerraQ (Project-ID 434617780 – SFB 1464), German Aerospace Center (DLR) – Q-BAGS (Project-ID 50WM2181), Germany's Excellence Strategy – EXC-2123 "QuantumFrontiers" – 390837967 and the European Union – CARIOQA-PMP (Project-ID 101081775). Computations were carried out using resources of the cluster system at the Leibniz University of Hannover, funded by the Deutsche Forschungsgemeinschaft (DFG) – INST 187/742-1 FUGG.

### Author Contributions

Conceptualization, methodology, and data curation were performed by M.R. Funding acquisition was secured by M.R. and J.M. Formal analysis was conducted by M.R. and J.M. Investigation was carried out by M.R., S.E., and A.K. Project administration was managed by M.R. and J.M. Software was developed by M.R. and S.E. Supervision was provided by J.M. and M.R. Visualization was prepared by M.R., S.E., and A.K. The original draft was written by M.R., S.E., and A.K. Review and editing were contributed by M.R., J.M., S.E., A.K., A.Kn., N.F., and M.S.

**Conflict of interest**

The authors declare that they have no known competing financial interests or personal relationships that could have appeared to influence the work reported in this paper.

**Data availability**

The datasets generated during and/or analyzed during the current study are available from the corresponding author on reasonable request.